# A Novel PN junction between Mechanically Exfoliated $\beta$-Ga$_2$O$_3$ and p-GaN

Jossue Montes[1], Chen Yang[1], Houqiang Fu[1], Tsung-Han Yang[1], Xuanqi Huang[1], Jingan Zhou[1], Hong Chen[1], Kai Fu[1], and Yuji Zhao[1,*]

[1]School of Electrical, Computer and Energy Engineering, Arizona State University, Tempe, AZ 85287, USA
[*]Email: yuji.zhao@asu.edu

Abstract

Several pn junctions were constructed from mechanically exfoliated ultrawide bandgap (UWBG) beta-phase gallium oxide ($\beta$-Ga$_2$O$_3$) and p-type gallium nitride (GaN). The mechanical exfoliation process, which is described in detail, is similar to that of graphene and other 2D materials. Atomic force microscopy (AFM) scans of the exfoliated $\beta$-Ga$_2$O$_3$ flakes show very smooth surfaces with average roughness of 0.647 nm and transmission electron microscopy (TEM) scans reveal flat, clean interfaces between the $\beta$-Ga$_2$O$_3$ flakes and p-GaN. The device showed a rectification ratio around 541.3 ($V_{+5}/V_{-5}$). Diode performance improved over the temperature range of 25°C and 200°C, leading to an unintentional donor activation energy of 135 meV. As the thickness of exfoliated $\beta$-Ga2O3 increases, ideality factors decrease as do the diode turn on voltages, tending toward an ideal threshold voltage of 3.2 V as determined by simulation. This investigation can help increase study of novel devices between mechanically exfoliated $\beta$-Ga$_2$O$_3$ and other materials.

Introduction

Beta-phase gallium oxide ($\beta$-Ga$_2$O$_3$) is a highly stable ultrawide bandgap (UWBG) semiconductor with a 4.6-4.9 eV bandgap [1, 2], which enables it to be used in deep ultra violet (DUV) and ultra-high power [3] applications. With a very high theoretical breakdown electric field of 8 MV/cm, saturation electron velocity of $2\times10^7$ cm/s, and Baliga's Figure of Merit (BFOM) of 3214.1 [4-9], $\beta$-Ga$_2$O$_3$ shows tremendous potential to perform beyond current high-power semiconductors such as GaN and SiC. A remarkably high interest in research on the material has arisen from the ready availability of large, low-cost, high-quality wafers of bulk $\beta$-Ga$_2$O$_3$ [10] but also from its unique monoclinic crystal structure. $\beta$-Ga$_2$O$_3$ is a 3D crystal that belongs to the C2/m space group with lattice constants **a**=1.22 nm, **b**=0.303 nm, and **c**=0.580 nm, with angle $\beta$=103.83° [11]. Because its **a** lattice vector, which points in the [100] direction, is significantly larger than the other two, a mechanical 'peeling off' or 'cleaving' of layers of the $\beta$-Ga$_2$O$_3$ is possible. The layers may then be transferred to any arbitrary substrate, where they will remain well-adhered, surviving subsequent ultrasonic wet cleaning and photolithographic processes. This process is similar to that done on 2D materials like graphene (for which the 2010 Nobel prize was awarded [12]) and transition metal dichalcogenides (TMDs) such as MoS$_2$ and WSe$_2$ [13, 14,]. The peeling off layers is known as mechanical exfoliation (more commonly called the Scotch tape method) and has opened the prospect of a great number of novel devices constructed using mechanically exfoliated $\beta$-Ga$_2$O$_3$ [15-17] and other semiconductor materials.

Very recently, electronic and optoelectronic devices constructed from mechanically exfoliated $\beta$-Ga2O3 flakes have enjoyed considerable attention, including diodes [18], transistors [19-25], photodetectors [26-29]. Like most other UWBG materials [10], $\beta$-Ga2O3 also shows promise in harsh-environment applications i.e. operating under radiation hazards [30]. Due to an exceedingly high hexagonal symmetry match with the wide bandgap (WBG) semiconductor GaN, devices between β-Ga2O3 and epitaxial GaN are an appealing consideration [31]. We report construction of the very first PN heterojunction diode between mechanically exfoliated β-Ga$_2$O$_3$ and p-type GaN. The diode shows a high rectification ratio and the turn on voltage is comparable to our simulation conducted by Silvaco. Remarkably, the electrical properties, including the turn on voltage and ideality factor improved in a series of elevated temperatures from 25°C to 200°C. By controlling the peeling off process, we fabricated diode with β-Ga$_2$O$_3$ in different thicknesses, and the device characteristics with respect to β-Ga$_2$O$_3$ thickness were also analyzed.

Experimental details

The p-type GaN target substrate consists of 300 μm bulk GaN with 300 nm of p-GaN ($N_A = 10^{19}$ cm$^{-3}$). The p-GaN was cleaned in acetone and isopropyl alcohol (IPA) for 5 minutes each under ultrasonic agitation in order to remove any residual organic contamination on the surface. Metal stacks of Pd (30 nm) / Ni (20 nm) / Au (150 nm) were deposited using electron beam evaporation on the p-GaN and subsequently annealed under N$_2$ at 1000°C for 30 s to form the p contact. 2-inch wafers of ($\bar{2}$01)-oriented β-Ga$_2$O$_3$ (Sn doping, $N_D = 5 \times 10^{18}$ cm$^{-3}$) were purchased from Tamura Corporation with bulk thickness 0.65±0.02 mm. These wafers were diced so as to expose the (100) plane and a thick fragment of β-Ga$_2$O$_3$ (~100 μm) was removed (Fig. 1 (a)). Using electron beam evaporation, metal stacks of Ti (20 nm) / Al (30 nm) / Ni (20 nm) / Au (100 nm) were deposited on the cleaved β-Ga$_2$O$_3$ fragment to form the n contact (Fig. 1 (b)). The contact was subsequently annealed in N$_2$ at 470°C for 1 minute (Fig. 1 (c)). The metal-semiconductor stack was placed upside down on Revalpha Heat Release tape (#3193MS) with the β-Ga$_2$O$_3$ exposed (Fig. 1 (d)). Regular Scotch tape was used to peel off layers of β-Ga$_2$O$_3$ from the bulk, reducing the thickness over many exfoliations (Fig. 1 (e)). After the final exfoliation with Scotch tape, the β-Ga$_2$O$_3$, metal, and thermal tape stack was then immediately transferred to the p-GaN substrate (Fig. 1 (f)). The backside of the tape was pressed down firmly on the p-GaN substrate. At this step, the β-Ga$_2$O$_3$ will adhere on to the p-GaN by its own pseudo-Van der Waals attractive forces – no adhesive material is required. The entire structure was placed upside down on a hot plate set to 120°C with vacuum seal beneath (Fig. 1 (g)). The upside down placement and vacuum seal are to ensure that the heat is evenly distributed across the thermal tape so that the adhesion strength of the tape evenly decreases across the area of the tape. After 2 minutes on the hot plate, the adhesive strength of the thermal tape vanishes completely, and the stack may be separated cleanly from the tape. Mechanical exfoliation by this method will result in β-Ga$_2$O$_3$ flakes ranging between 100 nm to over 100 μm thickness, depending on the number of times peeling off is performed.

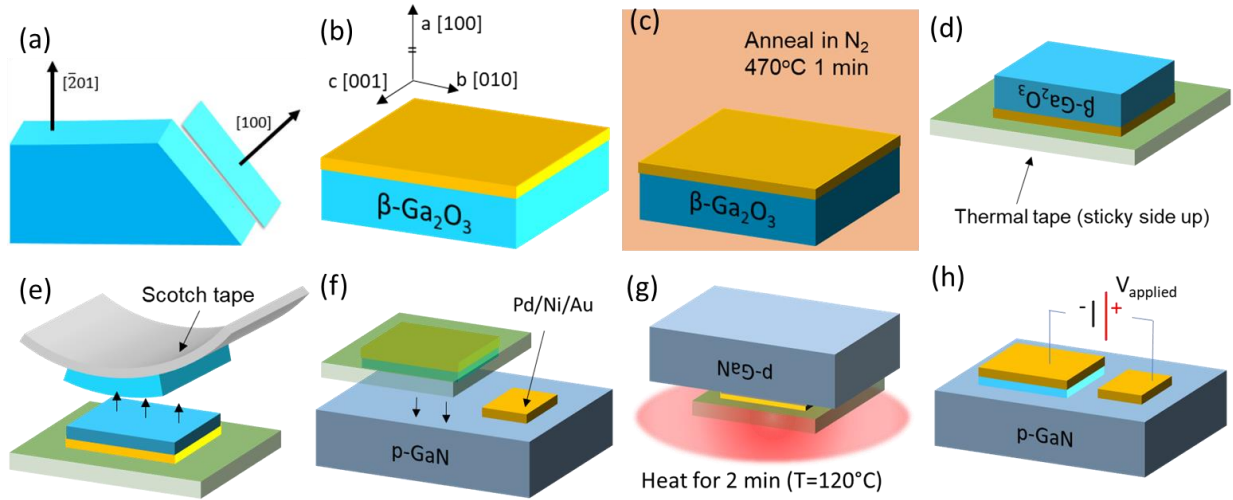

Fig. 1. Mechanical exfoliation of β-Ga$_2$O$_3$. (a) The bulk ($\bar{2}$01) wafers of β-Ga$_2$O$_3$ can be cleaved to expose the (100) plane. (b) Metal deposition via electron beam evaporation to deposit the n contact on the β-Ga$_2$O$_3$. (c) Anneal the contact in high-purity N$_2$ at 470°C for 1 minute. (d) Thermal tape is placed over the β-Ga$_2$O$_3$, metal stack and turned upside down. (e) Ordinary scotch tape is placed sticky side-down over the exposed β-Ga$_2$O$_3$ and peeled off, removing layers of the β-Ga$_2$O$_3$. (f) The β-Ga$_2$O$_3$, metal, thermal tape is placed on p-type GaN, which had had a p-contact deposited beforehand. (g) The entire stack is placed upside down on a vacuum-sealed hot plate at 120°C for 2 minutes to evenly distribute the heat across the thermal tape. (h) The finished device is removed and ready for testing.

Results and Discussions

Atomic force microscopy (AFM) scans were performed using a Bruker Multimode instrument to examine the surface roughness of the exfoliated flakes (see Fig. 2(a) and Fig. 2(b)). Using a 5×5 μm square area for each scan size, several scans showed an average surface roughness of 0.647 nm, comparable to previous studies on mechanically exfoliated β-Ga$_2$O$_3$ [21]. High resolution x-ray diffraction (HR-XRD) scans were performed on the bulk (as-received) β-Ga$_2$O$_3$ to examine the crystal quality.. The scans were carried out with a PANalytical X'Pert Pro diffractometer using Cu Kα1 radiation, a hybrid monochromator for the incident beam optics and a triple axis module for diffracted beam optics. The full-width at half-maximum of the ($\bar{2}$01) and (010) planes is 87 arcsec and 90 arcsec, respectively [8]. Fig. 2(c) shows the height profile of a 20 μm exfoliated flake (inset: optical microscope image of the flake). Transmission electron microscopy was used to investigate the interfaces of the device. A cross-sectional scan, prepared using focused ion beam (FIB), is presented in Fig. 2(d), showing the thickness of the layers. The β-Ga$_2$O$_3$ layer is flat, consistent with its characterization as a pseudo-2D material.

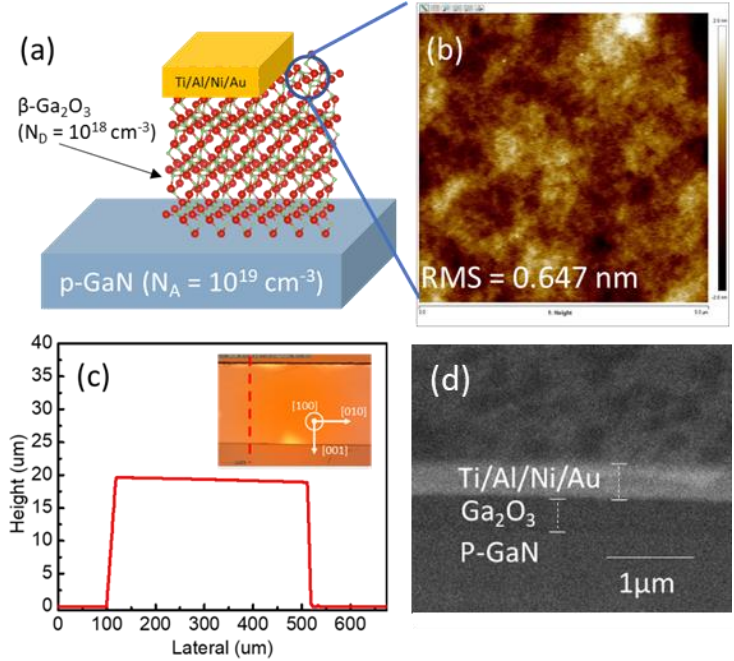

Fig. 2 Characterizations of the exfoliated β-Ga2O3 flake. (a) Schematic of the exfoliated β-$Ga_2O_3$ sample. (b) AFM image of the exfoliated β-$Ga_2O_3$ flake (without metal). (c) Height profile of a mechanically exfoliated flake (inset: optical microscope image). No metal is present on this flake. (d) TEM cross-section image of the exfoliated β-$Ga_2O_3$ flake.

A typical I-V characteristic of the heterojunction diode by a Keithley 2410 source meter is shown in Fig. 3(a). When a forward bias was applied to the heterojunction, the current increased to a large value (approximate 1.2 mA at 10 V). The turn-on voltage is defined as the voltage value at which significant current ($1\times10^{-5}$A) begins to flow. However, at reverse bias, the device showed a rectification property with a rectification ratio around 541.3 ($V_{+5}/V_{-5}$), as shown in Fig. 3(a) inset. To rule out the possibility of having a Schottky diode, which could be formed by the electrode metal on the β-$Ga_2O_3$ flake accidently contacting with the p-GaN, the I-V characteristic of one probe directly placed on p-GaN, which formed a Schottky barrier, was tested. The comparison curve in Fig. 3(b) and its inset shows a drastic difference between the heterojunction and the Schottky barrier, with huge disparity in turn-on voltage and reverse leakage. To further verify the performance of the diode, commercial software Silvaco was used to simulate the energy band alignment for the pn junction. The p-GaN was heavily doped ($N_A = 10^{19}$ cm$^{-3}$), and the β-$Ga_2O_3$ concentration was $5\times10^{18}$ cm$^{-3}$, according to Tamura Corporation. The band structure showed a valence and conduction band offsets of 1.27 and 0.11 eV, respectively; these values highly matched a related work [32]. Several of the diodes were fabricated and tested, and their turn-on voltages were comparable to the calculated ideal turn on voltage, thus confirming that the exfoliated β-$Ga_2O_3$ formed a heterojunction diode with p-GaN.

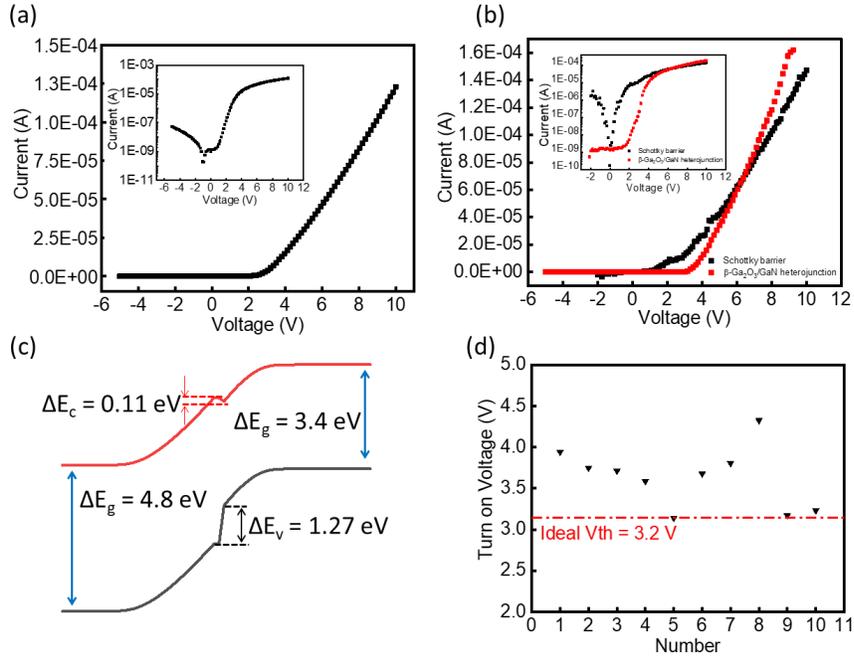

Fig. 3. (a) A typical I-V characteristic of the exfoliated n-β-Ga$_2$O$_3$/p-GaN heterojunction (inset: logarithmic scale). (b) Band structure simulated by Silvaco. (c) Comparison of the PN junction with a Schottky barrier diode (inset: logarithmic scale). (d) Turn-on voltage in a series of samples. The voltage when the forward current reach $10^{-5}$ A.

The temperature-dependent electrical tests were conducted in the range of 25°C to 200°C (Fig. 4(a)). The ideality factor and turn on voltage of the diode can be seen decreasing with increasing temperature (inset of Fig. 4(a)). This increased electrical behavior is due to the improvement in electrical conductance of the β-Ga$_2$O$_3$, which was also observed in the literature [33]. The conductance was given by differential conductance at 3.5 V forward bias. The activation energy estimated from Arrhenius plot in Fig. 4(b) is 135 meV over the temperature range considered, implying that improvement of n-type conductivity in β-Ga$_2$O$_3$ is due to the activation of unintentional deep donors [34], namely oxygen vacancies and other impurities [35, 36].

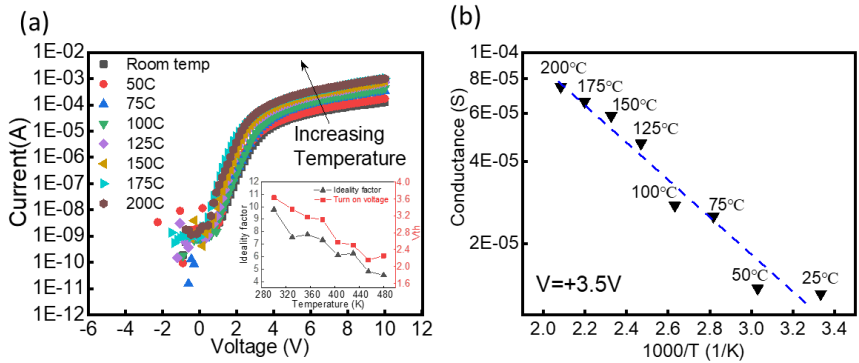

Fig. 4 Performance of the p-GaN/n-β-Ga$_2$O$_3$ diode with respect to temperature. (a) High-temperature I-V characteristic (inset: turn on voltage and ideality factor). (b) Electrical conductance at +3.5 V range from 25°C to 200°C.

The electrical characteristics of the diodes as a function of different β-Ga$_2$O$_3$ thicknesses were also studied. For the 100 nm, 5 μm, and 20 μm diodes, the I-V characteristics are shown in Fig. 5(a). The turn-on voltages are 4.09 V, 3.7 V, and 3.3 V and the ideality factors are 11.51, 8.34, and 5.91, respectively. An improvement in electrical performance with respect to thickness is shown in Fig. 5(b). The reason is likely that in order to produce thinner β-Ga$_2$O$_3$ flakes, more peeling off by scotch tape needs to be performed (Fig. 1(e)). The crystal quality of the β-Ga$_2$O$_3$ flake could be impacted during each application. Besides, tape residue may also accumulate at the interface between β-Ga$_2$O$_3$ and p-GaN, which could be detrimental to the electrical characteristics of diode. Using Inductively Coupled Plasma Etching (ICP) fluorine-based etching for β-Ga$_2$O$_3$ thinning and removal of tape residue may improve the diode performance [15].

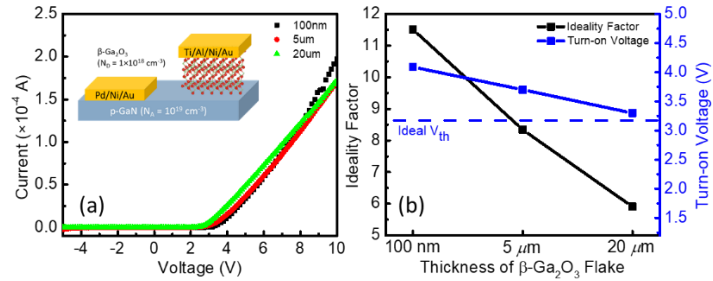

Fig. 5 Electrical characteristics of diodes with different β-Ga$_2$O$_3$ thicknesses. (a) Forward-bias I-V curve. (b) Ideality factors and turn-on voltage.

Conclusion

We have demonstrated the first pn heterojunction diode constructed between mechanically exfoliated β-Ga$_2$O$_3$ and p-GaN. The mechanical exfoliation process was described in detail. The electrical character of the pn junctions were tested with I-V and temperature measurements, and exfoliated β-Ga$_2$O$_3$ flakes measuring 100 nm, 5 μm, and 20 μm were compared. It was observed that the electrical properties of the diodes become better as temperature increased up to 200°C. As the thicknesses of the β-Ga$_2$O$_3$ flakes increased, the device performance also improved. The unintentional donor activation energy was calculated as 135 meV in the temperature range studied. At the max thickness of 20 μm, the pn junction tends toward the ideal turn-on voltage of 3.2 V, as determined by simulation. This work shows the potential of incorporating 2D β-Ga$_2$O$_3$ in GaN high power and high temperature devices.

This work was supported by the ARPA-E PNDIODES Program monitored by Dr. Isik Kizilyalli and partially supported by the NASA HOTTech Program grant number 80NSSC17K0768. We acknowledge the use of facilities within the Eyring Materials Center at Arizona State University. The device fabrication was performed at the Center for Solid State Electronics Research at Arizona State University. Access to the NanoFab was supported, in part, by NSF Contract No. ECCS-1542160.